\documentstyle[12pt]{article}
\title{Mean Field and Collisional Dynamics of Interacting
Fermion-Boson Systems in a Soluble Model}

\author{E.R. Takano Natti\thanks{E-MAIL:  erica@fma.if.usp.br}$\;\;$
and A.F.R de Toledo Piza \\ \it Instituto de F\'{\i}sica, Universidade
de S\~ao Paulo,\\ \it C.P. 66318, 05389-970 S\~ao Paulo, SP, Brasil}
\begin{document}

\maketitle

\begin{abstract}

A general time-dependent projection technique is applied to the study
of the dynamics of quantum correlations in a system consisting of
interacting fermionic and bosonic subsystems, described by the
Jaynes-Cummings Hamiltonian. The amplitude modulation of the Rabi
oscillations which occur for a strong, coherent initial bosonic field
is obtained from the spin intrinsic depolarization resulting from
collisional corrections to the mean-field approximation.\\ \\ PACS
number(s): 03.70+k, 42.50.Md, 32.80-t\\
Keywords: Double mean-field approximation; Collisional dynamics; Time-dependent projection; Fermion-boson system.
\end{abstract}

\section{Introduction}

Even very simple open subsystems of closed quantum mechanical systems
can display very intricate dynamical behavior, described by an
effective, non-unitary time evolution law. The quantum state of each
such subsystem can in fact be described in terms of a reduced density
operator which will in general evolve non-unitarily on the account of
correlations and decoherence effects involving different subsystems
\cite{NePi86,Jack89,Omne92}. The non-unitary effects will manifest
themselves specifically through the dynamical evolution of the
eigenvalues of the subsystem reduced density matrices, so that
individual subsystems evolve in general in a non-isoentropic fashion.

A first mean-field-like approximation to this general picture consists
in assuming isoentropic subsystem evolution under effective,
time-dependent Hamiltonian operators for each subsystem. This
approximation can be made self-consistent by allowing for explicit
dependence of each subsystem effective Hamiltonian on the present
state of the other, external subsystem \cite{NePi86}. Even when such
selfconsistency requirement is abandoned by simply parametrizing the
externally induced time dependence of the relevant subsystem effective
Hamiltonian, one is still left in general with a nonlinear dynamical
problem which defies exact treatment in a many-body or quantum field
theoretical context. In this case, the subsystem time evolution can be
analysed under a further mean field approximation through a
variational principle by making a Gaussian {\it Ansatz} for the
subsystem density matrix \cite{Jack89}.

In this work we implement the fully self-consistent {\it double}
time-dependent mean field approximation to a simple model system
consisting of interacting fermionic and bosonic subsystems described
by the Jaynes-Cummings Hamiltonian \cite{JayCum63}. Furthermore, by
using the projection approach developed by Nemes and Toledo Piza
\cite{NePi86} we obtain and evaluate explicitly corrections to this
self consistent mean field description. These corrections appear in
the form of suitable memory integrals added to the mean-field
dynamical equations. The resulting dynamical equations acquire then
the structure of kinetic equations, with the memory integrals
performing as collision terms which eliminate the isoentropic
mean-field constraint. Relevance of the Jaynes-Cummings model in this
context stems also from the fact that, besides being soluble, it can be
seen as corresponding to the relativistic scalar plasma \cite{Kal} in
zero spatial dimensions. The study of soluble models provides a clear
understanding of the physical phenomena involved and is useful in
controlling various approximations necessary for the treatment of more
realistic cases.

The definition of the relevant variables and our approach for studying
the dynamics of these variables are described in section 2. In section
3 we obtain the dynamical equations for these variables in the
collisional approximation . Finally we show and discuss some numerical
results, comparing to the exact solution of the model.

\section{Gaussian Variables and Their Effective Dynamics}

The Jaynes-Cummings Model (JCM) \cite{JayCum63} describes the simplest
fully quantized version of a system consisting of two interacting
quantum subsystems of different nature: a two level atom (fermionic
system) and a quantized field mode (bosonic system).  The
corresponding exactly soluble Hamiltonian is given by

\begin{equation}
\label{1}
H=\frac{\epsilon}{2}(a^{\dag}_{1}a_{1}-a^{\dag}_{-1}a_{-1})+\omega
b^{\dag}b+\lambda(a^{\dag}_{1}a_{-1}b+a^{\dag}_{-1}a_{1}
b^{\dag})\;\;,
\end{equation}

\noindent
where the annihilation and creation operators $b$ and $b^{\dag}$
satisfy the boson commutation relation $[b,b^{\dag}]=1$ and the
fermion operators $a_{\lambda}$ and $a^{\dag}_{\lambda}$ satisfy the
anticommutation relations $\{a_{\lambda}, a_{\lambda'}^{\dag}\}=
\delta_{\lambda',\lambda}$. The solubility of the model stems from the
fact that the interaction term of $H$ is written in the so called
Rotating Wave Approximation, which consists in ignoring
``anti-ressonant'' terms involving $a^{\dag}_{1}a_{-1}b^{\dag}$ and its
Hermitian conjugate.

The general idea of our approach is to focus on the time evolution of
variables which are fully determined by the mean values of the
annihilation/creation operators and their bilinear or quadratic
combinations, henceforth refered to as gaussian observables.  The
state of the system is generally given in terms of a density matrix
${\cal F}$ in the Heisenberg picture. ${\cal F}$ is hermitian, time
independent and has unit trace.

The mean values of the boson operators are given by

\[
\langle b(t) \rangle\equiv {\cal B}_{t}=Tr_{\mbox{\tiny BF}}\;b(t)
{\cal F}
\]

\noindent and its complex conjugate. Here and in what follows the
symbol $Tr_{\mbox{\tiny BF}}$ denotes a trace over both bosonic and fermionic
variables. Partial traces over bosonic or fermionic variables will be
written as $Tr_{\mbox{\tiny B}}$ and $Tr_{\mbox{\tiny F}}$ respectively. Using them
we can define the shifted boson operators

\begin{equation}
\label{2}
d(t)\equiv b(t)-{\cal B}_{t}
\end{equation}

\noindent which have vanishing ${\cal F}$-expectation values.

Next, the mean values of the bilinear forms of boson operators can be
combined in an extended one boson plus pairing density matrix ${\cal
R}$ \cite{RiSch,Lin90}

\begin {eqnarray}
\label{3}
{\cal R}_{\mbox{\tiny B}}&=&\left[
\begin{array}{cc}
R_{\mbox{\tiny B}}    &    \Pi_{\mbox{\tiny B}}\\
                      &                        \\
\Pi_{\mbox{\tiny B}}^{*} & 1+R_{\mbox{\tiny B}}
\end{array}
\right]=\left[
\begin{array}{cc}
\langle d^{\dag}(t)d(t)\rangle    &
\langle d(t)d(t)\rangle   \\
                   &                     \\
\langle d^{\dag}(t)d^{\dag}(t)\rangle   &
\langle d(t)d^{\dag}(t)\rangle
\end{array}
\right]\\ \nonumber
\end{eqnarray}

\noindent
where the one-boson density matrix $R_{\mbox{\tiny B}}$ is hermitian
and the pairing density matrix $\Pi_{\mbox{\tiny B}}$ is symmetric.

For the fermionic system the extended density matrix is

\begin{eqnarray}
\label{4}
{\cal R}_{\mbox {\tiny F}}&=&\left[
\begin{array}{cc}
R_{\mbox{\tiny F}}        &  \Pi_{\mbox{\tiny F}} \\
             &          \\
-\Pi^{*}_{\mbox{\tiny F}} &  1-R_{\mbox{\tiny F}} 
\end{array}
\right]=\left[
\begin{array}{cc}
\langle a_{\lambda'}^{\dag}(t)a_{\lambda}(t)\rangle  &  
\langle a_{\lambda'}(t)a_{\lambda}(t)\rangle \\
                    &                 \\
\langle a_{\lambda'}^{\dag}(t)a_{\lambda}^{\dag}(t)\rangle  &
\langle a_{\lambda'}(t)a_{\lambda}^{\dag}(t)\rangle
\end{array}
\right] \\ \nonumber
\end{eqnarray}
 
\noindent
where the hermitian matrix $R_{\mbox{\tiny F}}$ and the antisymmetric
matrix $\Pi_{\mbox{\tiny B}}$ are the one-fermion and pairing density
matrices, respectively.

To deal with the pairing densities one defines new Bogoliubov
quasi-particle operators. In the case of the bosons we define
quasi-boson operators as

\begin{eqnarray}
\beta(t) &=& x^{*}_{t}(b-{\cal B}_{t})+y^{*}_{t}(b^{+}-{\cal
B}^{*}_{t}) \nonumber\\ \label{5}\\ \beta^{+}(t)&=& x_{t}(b^{+}-{\cal
B}^{*}_{t})+y_{t}(b-{\cal B}_{t}) \nonumber
\end{eqnarray}  

\noindent and require that $\langle \beta\beta\rangle=\langle
\beta^{\dag}\beta^{\dag}\rangle=0$. The preservation of the
commutation relations requires furthermore that the transformation
coefficients $x_{t}$ and $y_{t}$ be chosen so that
$|x_{t}|^{2}-|y_{t}|^{2}=1$. Thus the densities $\langle
b^{\dag}b^{\dag}\rangle$ and $\langle bb\rangle$ can be parametrized
in terms of $x_{t}$, $y_{t}$ and $\nu=\langle
\beta^{\dag}\beta\rangle$.

The coefficients $x_{t}$ and $y_{t}$ of the Bogoliubov transformation
are determined by solving the secular problem

\begin{equation}
\label{6}
G{\cal R}_{\mbox{\tiny B}}X_{\mbox{\tiny B}}=X_{\mbox{\tiny
B}}GQ_{\mbox{\tiny B}}
\end{equation}

\noindent where

\begin{equation}
\label{7}
X_{\mbox{\tiny B}}=\left[
\begin{array}{cc}
x_{t}   &   y_{t}^{*}  \\
        &             \\
y_{t}   &   x_{t}^{*}
\end{array}
\right]\;\;\; G=\left[
\begin{array}{cc}
1  &  0  \\
   &     \\
0  & -1  
\end{array}
\right]\;\;\;Q_{\mbox{\tiny B}}=\left[
\begin{array}{cc}
\nu  &    0    \\
     &         \\
 0   &  1+\nu
\end{array}
\right]\;\;.
\end{equation}

\noindent
The eigenvalue $\nu$ can be interpreted as shifted boson occupation
number for the paired natural orbital described by $X_{\mbox{\tiny
B}}$.  Since the Bogoliubov transformation is canonical one can verify
that $X_{\mbox{\tiny B}}$ satisfies the orthogonality and completeness
relation

\begin{equation}
\label{8}
X_{\mbox{\tiny B}}^{\dag}GX_{\mbox{\tiny B}}=X_{\mbox{\tiny
B}}GX_{\mbox{\tiny B}}^{\dag}=G\;\;.
\end{equation}

For the fermion operators we define the transformation

\begin{equation}
\label{9}
\left[
\begin{array}{c}
\alpha_{1}(t)\\
             \\
\alpha_{-1}(t)\\
              \\
\alpha^{\dag}_{1}(t)\\
                 \\
\alpha^{\dag}_{-1}(t)
\end{array}
\right]=
\left[
\begin{array}{cc}
X^{*}  &  0   \\
       &      \\
0      &  X         
\end{array}
\right]
\left[
\begin{array}{c}
a_{1}(t)\\
        \\
a_{-1}(t)\\
         \\
a^{\dag}_{1}(t)\\
               \\
a^{\dag}_{-1}(t)
\end{array}
\right]
\end{equation}

\noindent where

\begin{equation}
\label{10}
X=\left[
\begin{array}{cc}
u_{t}   &   -v_{t}   \\
        &            \\
v_{t}^{*} & u_{t}^{*}
\end{array}
\right]\;\;.
\end{equation}

\noindent
Here we require that $\langle
\alpha_{\lambda'}(t)\alpha_{\lambda}(t)\rangle=\langle
\alpha_{\lambda'}^{\dag}(t)\alpha_{\lambda}^{\dag}(t)\rangle=0 $ and
the fermion densities $\langle
a_{\lambda'}^{\dag}(t)a_{\lambda}(t)\rangle$ and $\langle
a_{\lambda'}(t)a_{\lambda}^{\dag}(t)\rangle$ are parametrized in terms
of $u_{t}$, $v_{t}$ and $p_{\lambda}=\langle
\alpha_{\lambda}^{\dag}\alpha_{\lambda}\rangle,\;\lambda=-1,1$.  As in
the case of the bosons we have an eigenvalue problem

\begin{equation}
\label{11}
{\cal X}_{\mbox{\tiny F}}^{\dag}{\cal R}_{\mbox{\tiny F}}{\cal
X}_{\mbox{\tiny F}}=Q_{\mbox{\tiny F}}
\end{equation}
\vskip 0.3cm

\noindent where ${\cal X}_{\mbox{\tiny F}}$ is the transformation
defined in (\ref{9})

\begin{equation}
\label{12}
{\cal X}_{\mbox{\tiny F}}=\left[
\begin{array}{cc}
X^{*}  &  0  \\
       &     \\
0      &  X  
\end{array}
\right]\;\;\mbox{and}\;\;{\cal X}_{\mbox{\tiny F}}^{\dag}=\left[
\begin{array}{cc}
X^{T}  &  0  \\
       &     \\
0      &  X^{\dag}
\end{array}
\right]
\end{equation}
\vskip 0.3cm

\noindent satisfying the unitary condition

\begin{equation}
\label{13}
{\cal X}_{\mbox{\tiny F}}^{\dag}{\cal X}_{\mbox{\tiny F}}={\cal
X}_{\mbox{\tiny F}}{\cal X}_{\mbox{\tiny F}}^{\dag}={\bf I}_{4}.
\end{equation}
\vskip 0.3cm

\noindent The matrix $Q_{\mbox{\tiny F}}$ is the extended density
matrix in quasi-particle basis

\begin{equation}
\label{14}
Q_{\mbox{\tiny F}}=\left[
\begin{array}{cc}
{\bf p}_{\lambda}  &      0        \\
                   &               \\
     0             & 1-{\bf p}_{\lambda}
\end{array}
\right]=\left[
\begin{array}{cc}
\langle \alpha^{\dag}_{\lambda'}(t)\alpha_{\lambda}(t)\rangle   & 
0 \\
                          &                        \\
0 &
\langle \alpha_{\lambda'}(t)\alpha^{\dag}_{\lambda}(t)\rangle
\end{array}
\right]
\end{equation}
\vskip 0.3cm

\noindent where ${\bf p}_{\lambda}$ is the one-fermion density matrix
in quasi-particle basis. In this basis ${\bf p}_{\lambda}$ is diagonal
with eigenvalues $p_{1}$ and $p_{-1}$ which correspond to the
occupation numbers of the natural orbitals.

The next step is to obtain the time evolution of the mean values of
gaussian observables.  We begin with the boson subsystem. For ${\cal
B}_{t}$ one finds immediately

\begin{equation}
\label{15}
i\dot {\cal B}_{t}=Tr_{\mbox{\tiny BF}}[b,H]{\cal F}=x_{t}
Tr_{\mbox{\tiny BF}}[\beta,H]{\cal F}-y_{t}^{*}Tr_{\mbox{\tiny
BF}}[\beta^{\dag},H]{\cal F}
\end{equation}
\vskip 0.5cm

\noindent where $H$ is the Hamiltonian given by (\ref{1}). For the
remaining quantities, we first rewrite the eigenvalue equation
(\ref{6}), using (\ref{8}), as

\begin{equation}
\label{16}
X^{\dag}{_{\mbox{\tiny B}}\cal R}_{\mbox{\tiny B}}X_{\mbox{\tiny B}}=
Q_{\mbox{\tiny B}}\;\;.
\end{equation}

\noindent Taking the time derivative we have

\begin{eqnarray}
\label{17}
X^{\dag}_{\mbox{\tiny B}}\dot{\cal R}_{\mbox{\tiny B}}X_{\mbox{\tiny
B}}&=&\dot Q_{\mbox{\tiny B}}-\dot X^{\dag}_{\mbox{\tiny B}}{\cal
R}_{\mbox{\tiny B}}X_{\mbox{\tiny B}}-X^{\dag}_{\mbox{\tiny B}}{\cal
R}_{\mbox{\tiny B}}\dot X_{\mbox{\tiny B}}\nonumber\\ &=&\dot
Q_{\mbox{\tiny B}}-\dot X^{\dag}_{\mbox{\tiny B}}GX_{\mbox{\tiny B}}G
Q_{\mbox{\tiny B}}-Q_{\mbox{\tiny B}}GX^{\dag}_{\mbox{\tiny B}}G\dot
X_{\mbox{\tiny B}}\;\;.
\end{eqnarray}
\vskip 0.3cm
 
The left hand side of this equation can be evaluated using the
Heisenberg equation of motion

\begin{equation}
\label{18}
iX_{\mbox{\tiny B}}^{\dag}\dot {\cal R}_{\mbox{\tiny B}}X_{\mbox{\tiny
B}}=\left(
\begin{array}{cc}
Tr_{\mbox{\tiny BF}}[\beta^{\dag}\beta,H]{\cal F} & Tr_{\mbox{\tiny
                 BF}}[\beta\beta,H]{\cal F} \\ & \\ Tr_{\mbox{\tiny
                 BF}}[\beta^{\dag}\beta^{\dag},H]{\cal F} &
                 Tr_{\mbox{\tiny BF}}[\beta\beta^{\dag},H]{\cal F}
\end{array}
\right)\;\;.
\end{equation}
\vskip 0.5cm

\noindent The right hand side of eq. (\ref{17}), on the other hand can
be rewritten using (\ref{7}). Equating the result to (\ref{18}) we
have equations which describe the time evolution of the bosonic
variables explicitly :

\begin{eqnarray}
\label{19}
i\dot \nu &=& Tr_{\mbox{\tiny BF}}[\beta^{\dag}\beta,H] {\cal F} \\
\nonumber\\
\label{20}
i(1+2\nu)(\dot x_{t}^{*}y_{t}^{*}-x_{t}^{*}\dot y_{t}^{*})&=&
Tr_{\mbox{\tiny BF}}[\beta\beta,H] {\cal F}.\\ \nonumber
\end{eqnarray}
 
We have next to obtain analogous equations for the fermionic
variables. Taking the time derivative of the eigenvalue equation
(\ref{11}) and using the unitary condition (\ref{13}) we obtain

\begin{equation}
\label{21}
{\cal X}_{\mbox{\tiny F}}\dot {\cal R}_{\mbox{\tiny F}}{\cal
X}_{\mbox{\tiny F}}=\dot Q_{\mbox{\tiny F}}-\dot {\cal X}_{\mbox{\tiny
F}}^{\dag}{\cal X}_{\mbox{\tiny F}}Q_{\mbox{\tiny F}}-Q_{\mbox{\tiny
F}}{\cal X}_{\mbox{\tiny F}}^{\dag}\dot {\cal X}_{\mbox{\tiny F}}\;\;.
\end{equation}
\vskip 0.3cm

\noindent Substituting (\ref{12}) and (\ref{14}) and using the
unitary condition (\ref{13}) we can evaluate the right hand side of
(\ref{21}). To evaluate the left hand side we use the Heisenberg
equation of motion. We obtain

\begin{eqnarray}
\label{22}
i\dot p_{\lambda}&=&Tr_{\mbox{\tiny
BF}}\left[\alpha_{\lambda}^{\dag}\alpha_{\lambda},H\right] {\cal
F}\;\;,\;\;\;\;\lambda=-1,1 \\\nonumber\\
\label{23}
i(p_{1}-p_{-1})(\dot u_{t}v_{t}-\dot
v_{t}u_{t})&=&Tr_{\mbox{\tiny
BF}}\left[\alpha_{1}^{\dag}\alpha_{-1},H\right] {\cal F}\\ \nonumber
\end{eqnarray}

\noindent which describe the dynamics of the fermionic variables.

Eqs. (\ref{15}), (\ref{19}), (\ref{20}), (\ref{22}) and (\ref{23}) are
however not closed, ${\cal F}$ is not fully determined by gaussian
variables. In order to deal with this situation we split the full
density as

\begin{eqnarray}
\label{30}
{\cal F}&=&{\cal F}_{0}(t)+{\cal F}'(t)\nonumber\\ &\equiv& {\cal
F}^{\mbox{\tiny B}}_{0}{\cal F}^{\mbox{\tiny F}}_{0}+{\cal F}'(t)
\end{eqnarray}

\noindent where the factorized form of ${\cal F}_{0}(t)$ embodies what
we refer to as the double mean field approximation. The subsystem
densities ${\cal F}^{\mbox{\tiny B}}_{0}$ and ${\cal F}^{\mbox{\tiny
F}}_{0}$ are in fact gaussian densities, written in the the form of an
exponential of a  bilinear, Hermitian expression in the
creation/annihilation parts of the bosonic and of the fermionic fields
respectively. When explicitly normalized to unit trace they
are given in terms of the transformed operators introduced in
Eqs. (\ref{5}) and (\ref{9}) as \cite {Lin90,Cloise}

\begin{eqnarray}
\label{31}
{\cal F}^{\mbox{\tiny F}}_{0}&=&\prod_{\lambda}
[p_{\lambda}\alpha_{\lambda}^{\dag}\alpha_{\lambda}+(1-p_{\lambda})\alpha_{\lambda}\alpha_{\lambda}^{\dag}]\\
\nonumber \\
\label{32}
{\cal F}^{\mbox{\tiny B}}_{0}&=&\frac {1}{1+\nu}\left(\frac
{\nu}{1+\nu}\right)^{\beta^{\dag}\beta}.
\end{eqnarray}

\noindent The remainder density ${\cal F}'(t)$ is consequently a
traceless, pure correlation part of the full density. In view of the
special form used for ${\cal F}_{0}(t)$ it will in general contain
correlations of two types : {\it inter-subsystem} (boson-fermion)
correlations and {\it intra-subsystem} (boson-boson and
fermion-fermion) correlations.

The next step consists in using the fact that the decomposition
(\ref{30}) can be implemented in terms of a time-dependent projection
operator ${\bf\cal P}(t)={\bf\cal P}(t){\bf\cal P}(t)$
(cf. Ref. \cite{WiPi74})

\begin{equation}
\label{33}
{\cal P}\cdot = {\cal F}^{\mbox{\tiny F}}_{0}Tr_{\mbox{\tiny F}}{\cal
P}_{\mbox{\tiny B}}\cdot +{\cal F}^{\mbox{\tiny B}}_{0}Tr_{\mbox{\tiny
B}}{\cal P}_{\mbox{\tiny F}}\cdot -{\cal F}^{\mbox{\tiny F}}_{0}{\cal
F}^{\mbox{\tiny B}}_{0}Tr_{\mbox{\tiny F}}Tr_{\mbox{\tiny B}}\cdot
\end{equation}
\vskip 0.5cm

\noindent where ${\cal P}_{\mbox{\tiny B}}$ is the boson projection
operator \cite{Lin}

\begin{eqnarray}
\label{34}
{\cal P}_{\mbox{\tiny B}}\cdot&=& \left[\left(1-\frac
{\beta^{\dag}\beta-\nu}{1+\nu}\right)Tr_{\mbox{\tiny B}}\cdot+\frac
{\beta^{\dag}\beta -\nu}{\nu(1+\nu)}Tr_{\mbox{\tiny
B}}(\beta^{\dag}\beta \cdot)+\frac {\beta}{\nu}Tr_{\mbox{\tiny
B}}(\beta^{\dag}\cdot)+ \right.\nonumber\\ &+&\left.\frac
{\beta^{\dag}}{1+\nu}Tr_{\mbox{\tiny B}}(\beta \cdot)+\frac
{\beta\beta}{2\nu^{2}}Tr_{\mbox{\tiny
B}}(\beta^{\dag}\beta^{\dag}\cdot)+\frac
{\beta^{\dag}\beta^{\dag}}{2(1+\nu)^{2}}Tr_{\mbox{\tiny B}}(\beta\beta
\cdot)\right]{\cal F}^{\mbox{\tiny B}}_{0},
\end{eqnarray}

\noindent ${\cal P}_{\mbox{\tiny F}}$ is the fermion projector
\cite{BuFeNe88}

\begin{equation}
\label{35}
{\cal P}_{\mbox{\tiny F}}\cdot = \left[\left(1-\sum_{\lambda=-1,1}\frac
{\alpha_{\lambda}^{\dag}\alpha_{\lambda}-p_{\lambda}}{1-p_{\lambda}}\right)Tr_{\mbox{\tiny
F}}\cdot + \sum_{\lambda\prime,\lambda=-1,1}\left(\frac
{\alpha_{\lambda}^{\dag}\alpha_{\lambda\prime}-p_{\lambda}\delta_{\lambda\lambda\prime}}{p_{\lambda\prime}(1-p_{\lambda})}\right)
Tr_{\mbox{\tiny F}} (\alpha_{\lambda\prime}^{\dag}\alpha_{\lambda}\cdot)\right]{\cal
F}^{\mbox{\tiny F}}_{0}\;\;
\end{equation}

\noindent and the dot stands for the object on which the operator
acts. With the help of ${\bf\cal P}(t)$ the factorized gaussian
density can be obtained as 

\begin{equation}
\label{25}
{\cal F}_{0}(t)={\bf\cal P}(t){\cal F},
\end{equation}

\noindent while its time-derivative is given by the expression

\begin{equation}
\label{26}
i\dot  {\cal F}_{0}(t)=[ {\cal F}_{0}(t),H]+{\cal P}(t)[H, {\cal F}]
\end{equation}
\vskip 0.5cm

\noindent which is in fact the Heisenberg Picture counterpart of the
equation $\dot {\bf\cal P}{\cal F}=0$ which has been used to define
${\bf\cal P}(t)$ in Schr\"{o}dinger Picture \cite{WiPi74,BuFeNe88}.

The remaining formal steps towards closing the equations of motion for
the gaussian variables are now straightforward.  From (\ref{30}) and
(\ref{26}) we can obtain a differential equation for ${\cal F}'(t)$ in
terms of ${\cal F}_{0}(t)$

\begin{equation}
\label{27}
\left[i\frac {d}{dt} + {\cal P}(t){\cal L}\right]{\cal F}'(t) =
{\cal Q}{\cal L}{\cal F}_{0}(t)\;\;.
\end{equation}
\vskip 0.3cm

\noindent
where we introduced the complementary projector ${\cal Q}(t){\cal
F}\equiv({\cal I}-{\cal P}(t)){\cal F}={\cal F}'(t)$ and the
notation ${\cal L}\cdot$ for the Liouvillian time-displacement generator
$[H,\;\cdot\;]$.  This equation has the formal solution

\begin{equation}
\label{28}
{\cal F}'(t)={\cal G}(t,0){\cal F}'(0)-i\int_{0}^{t}
dt'\;{\cal G}(t,t'){\cal Q}(t'){\cal L}{\cal F}_{0}(t')\;\;,
\end{equation}
\vskip 0.3cm 

\noindent
where the first term accounts for initial conditions, and ${\cal
G}(t,t')$ is the time-ordered Green's function

\begin{equation}
\label{29}
{\cal G}(t,t')={\rm T} \exp\left[i\int _{t'}^{t} d\tau\; 
{\cal P}(\tau){\cal L}\right]\;\;.
\end{equation}
\vskip 0.5cm

\noindent Thus ${\cal F}'(t)$, and therefore ${\cal F}$, can be
formally expressed in terms of ${\cal F}_{0}(t')$ for $t'<t$ and of
{\it initial} correlations ${\cal F}'(0)$.  This finally allows one to
express the dynamical equations (\ref{15}), (\ref{19}), (\ref{20}),
(\ref{22}), and (\ref{23}) as traces over functionals of ${\cal
F}_{0}(t')$ and of the initial correlation part ${\cal F}'(0)$.

\section{Approximation to the Collision Dynamics}

As can be seen from the formally exact Eq. (\ref{28}), the correlation
contributions to the dynamical equations involve traces over
functionals of ${\cal F}_{0}(t')$ with memory effects besides the
initial correlations. Since these objects involve the complicated fully
correlated time-dependence of the Heisenberg operators, approximations
are clearly needed for their actual evaluation.

A systematic expansion scheme for the memory effects in which the mean
energy is conserved in all orders has been discussed in ref.
\cite{BuFeNe88} in terms of the Schr\"{o}dinger Picture. Following Lin
\cite{Lin} we implement in the Heisenberg picture a modified version
of the lowest order approximation given in that work which consists in
approximating the actual time evolution of the operators $\beta(t)$
when evaluating memory effects, by the effective mean field evolution

\begin{equation}
\label{36}
i\frac{d\beta}{dt}=[\beta,H_{0}]+i(\dot x_{t}^{*}x_{t}-\dot
y_{t}^{*}y_{t})\beta-i(\dot x_{t}^{*}y_{t}^{*}-x_{t}^{*}\dot
y_{t}^{*})\beta^{\dag} -Tr_{\mbox{\tiny BF}}[\beta,H]{\cal F}.
\end{equation}
\vskip 0.3cm

\noindent The last three terms account for the explicit time
dependence of the $\beta(t)$ related to the shift amplitudes ${\cal
B}_{t}$ and to the Bogoliubov transformation (\ref{5}).
For the fermion operators $\alpha_{1}(t)$ and $\alpha_{-1}(t)$ the
corresponding approximation reads

\begin{eqnarray}
i\frac{d\alpha_{1}}{dt}&=&[\alpha_{1},H_{0}]+i(\dot
u_{t}^{*}u_{t}+\dot v_{t}^{*}v_{t})\alpha_{1}+i(\dot
u_{t}^{*}v_{t}^{*}-u_{t}^{*}\dot v_{t}^{*})\alpha_{-1}\nonumber \\
\label{37}\\ i\frac{d\alpha_{-1}}{dt}&=&[\alpha_{-1},H_{0}]+i(\dot
u_{t}u_{t}^{*}+\dot v_{t}v_{t}^{*})\alpha_{-1}-i(\dot
u_{t}v_{t}-u_{t}\dot v_{t})\alpha_{1}. \nonumber \\ \nonumber
\end{eqnarray}

\noindent In these expressions the Hamiltonian written as $H_{0}(t)$ is
taken as the effective mean-field Hamiltonian
(cf. Ref. \cite{BuFeNe88})

\begin{eqnarray}
\label{38}
H_{0}&=&{\cal P}^{\dag}(t)H+\beta^{\dag}Tr_{\mbox{\tiny
BF}}[\beta,H]{\cal F}'(t)-\beta Tr_{\mbox{\tiny
BF}}[\beta^{\dag},H]{\cal F}'(t)\nonumber \\ \nonumber\\
&+&\frac{\beta^{\dag}\beta^{\dag}}{2(1+2\nu)}Tr_{\mbox{\tiny
BF}}[\beta\beta,H]{\cal
F}'(t)-\frac{\beta\beta}{2(1+2\nu)}Tr_{\mbox{\tiny
BF}}[\beta^{\dag}\beta^{\dag},H]{\cal F}'(t)\nonumber\\ \nonumber\\
&+&\frac{\alpha_{1}^{\dag}\alpha_{-1}}{(p_{-1}-p_{1})}Tr_{\mbox{\tiny
BF}}[\alpha_{-1}^{\dag}\alpha_{1},H]{\cal
F}'(t)+\frac{\alpha_{-1}^{\dag}\alpha_{1}}{(p_{1}-p_{-1})}Tr_{\mbox{\tiny
BF}}[\alpha_{1}^{\dag}\alpha_{-1},H]{\cal F}'(t)\;\;.\\ \nonumber
\end{eqnarray}

The first term in this expression correspond to the lowest
approximation according to ref. \cite{BuFeNe88}. Note that ${\cal
P}(t)$ is {\it not} an orthogonal projection, i.e. ${\cal
P}^\dagger(t) \neq {\cal P}(t)$. The remaining terms, included here,
represent correlation contributions to the effective mean-field.

Consistently with this approximation, the Green's function (\ref{29})
is in lowest order just the unit operator

\[
{\cal G}(t,t')={\cal I}
\]
 
\noindent so that the correlation density is approximated as

\begin{equation}
\label{39}
{\cal F}'(t)={\cal G}^{(0)}(t,0){\cal F}'(0)-i\int_{0}^{t}dt' {\cal
Q}(t'){\cal L}{\cal F}_{0}(t').
\end{equation}
\vskip 0.3cm

\noindent For simplicity, we restrict ourselves now to initially
uncorrelated states, i.e. ${\cal F}'(0)=0$. The dynamical equations of
the gaussian variables with correlation terms then become

\begin{eqnarray}
i\frac{d}{dt}\left[Tr_{\mbox{\tiny BF}}({\cal O}_{\mbox{\tiny F}}(t)
{\cal F})\right]&=&Tr_{\mbox{\tiny BF}}[{\cal O}_{\mbox{\tiny F}}(t)
,H]{\cal F}_{0}(t)-i Tr_{\mbox{\tiny BF}}[{\cal O}_{\mbox{\tiny
F}}(t),H] \int_{0}^{t}dt' {\cal Q}(t'){\cal L}{\cal
F}_{0}(t')\nonumber\\ & & \mbox{mean-field
approx.}\;\;\;\;\;\;\;\;\;\;\;\mbox{correlation term}\nonumber\\
\label{40}\\ i\frac{d}{dt}\left[Tr_{\mbox{\tiny BF}}({\cal
O}_{\mbox{\tiny B}}(t) {\cal F})\right]&=&Tr_{\mbox{\tiny BF}}[{\cal
O}_{\mbox{\tiny B}}(t), H]{\cal F}_{0}(t)-i Tr_{\mbox{\tiny BF}}[{\cal
O}_{\mbox{\tiny B}}(t),H]\int_{0}^{t}dt' {\cal Q}(t'){\cal L}{\cal
F}_{0}(t')\nonumber\\ & & \mbox{mean-field
approx.}\;\;\;\;\;\;\;\;\;\;\;\mbox{correlation term}\nonumber\\
\nonumber
\end{eqnarray}

\noindent where ${\cal O}_{\mbox{\tiny F}}$ and ${\cal O}_{\mbox{\tiny
B}}$ correspond to fermion and boson gaussian variables, respectively.

The terms involving ${\cal F}_{0}(t)$ can be calculated from $H$ in
quasi-particle basis and the projection operator ${\cal P}$ given by
(\ref{33}), (\ref{34}) and (\ref{35}).  However,the traces in the last
terms of (\ref{40}) still can not be taken directly, since the
quasi-particle operators in the integral and in the first commutator
are at different times. Therefore, we use the approximation in which
the time evolution of these operators is given by (\ref{36}) and
(\ref{37}). Using (\ref{38}), (\ref{20}) and (\ref{23}) we obtain

\begin{eqnarray}
\label{41}
i\frac{d\beta}{dt}&=&i(\dot x_{t}^{*}x_{t}-\dot
y_{t}^{*}y_{t})\beta+\omega(|x_{t}|^{2}+|y_{t}|^{2})\beta\equiv
F_{\beta}(t)\beta\\ \nonumber\\
\label{42}
i\frac{d\alpha_{1}}{dt}&\!=&\left[\frac{\epsilon}{2}
(|u_{t}|^{2}-|v_{t}|^{2})-\lambda(u_{t}^{*}v_{t}{\cal
B}_{t}+u_{t}v_{t}^{*}{\cal B}_{t}^{*})\right]\alpha_{1}+i(\dot
u_{t}^{*}u_{t}+\dot v_{t}^{*}v_{t})\alpha_{1}\equiv
F_{\alpha_{1}}(t)\alpha_{1}\\ \nonumber \\
\label{43}
i\frac{d\alpha_{-1}}{dt}&\!=&\!\!-\left[\frac{\epsilon}{2}
(|u_{t}|^{2}-|v_{t}|^{2})-\lambda(u_{t}^{*}v_{t}{\cal
B}_{t}+u_{t}v_{t}^{*}{\cal B}_{t}^{*})\right]\alpha_{-1}+i(\dot
u_{t}u_{t}^{*}+\dot v_{t}v_{t}^{*})\alpha_{-1}\equiv
F_{\alpha_{-1}}(t)\alpha_{-1}\nonumber\\
\end{eqnarray}

From (\ref{41}), (\ref{42}) and (\ref{43}) one can verify that the
operators $\beta(t)$, $\alpha(t)$ at different times are related as

\begin{equation}
\label{44}
i\frac {d\beta}{dt}\equiv F_{\beta}(t)\beta \;\;\Longrightarrow
\;\;\beta(t)=e^{i\varphi(t,t')}\beta(t')
\end{equation}
\vskip 0.5cm

\noindent where the phase $\varphi(t,t')$ is given by

\[
\varphi(t,t')=-\int_{t'}^{t} d\tau F_{\beta}(\tau)
\]

\noindent
and

\begin{equation}
\label{45}
i\frac{d\alpha_{1}}{dt}\equiv F_{\alpha_{1}}(t)\alpha_{1}\;\;
\Longrightarrow\;\;
\alpha_{1}(t)=e^{i\xi_{\alpha}(t,t')}\alpha_{1}(t')
\end{equation}

\begin{equation}
\label{46}
i\frac{d\alpha_{-1}}{dt}\equiv F_{\alpha_{-1}}(t)\alpha_{-1}
\;\;\Longrightarrow\;\;\alpha_{-1}(t)=e^{-i\xi_{\alpha}(t,t')}
\alpha_{-1}(t)
\end{equation}
\vskip 0.5cm

\noindent where the phase $\xi_{\alpha}(t,t')$ is 

\[
\xi_{\alpha}(t,t')=-\int_{t'}^{t} d\tau F_{\alpha}(\tau).
\]

Finally, the equations of motion that describe the present
approximation to the collisional dynamics are

\begin{eqnarray}
\label{47}
\dot\nu &=&\lambda^{2}\left\{(u_{t}^{*}v_{t}y_{t}^{*} -
u_{t}v_{t}^{*}x_{t}^{*})I_{1} + ({u_{t}^{*}}^{2}y_{t}^{*} +
{v_{t}^{*}}^{2}x_{t}^{*})I_{2} + (u_{t}^{2}x_{t}^{*} +
v_{t}^{2}y_{t}^{*})I_{3}\right\} +\nonumber\\
\nonumber\\ & +&\mbox{C.C.}\\ \nonumber\\
\label{48}
\frac{d}{dt}(p_{1}-p_{-1})&=&2\lambda^{2}\left\{-({u_{t}^{*}}^{2}
x_{t}+{v_{t}^{*}}^{2}y_{t})I_{3}^{*} + ({u_{t}^{*}}^{2}y_{t}^{*} +
{v_{t}^{*}}^{2}x_{t}^{*})I_{2}\right\}\nonumber\\
& +&\mbox{C.C.}\\ \nonumber\\
\label{49}
i(p_{1}-p_{-1})(\dot u_{t}v_{t}-u_{t}\dot
v_{t})&=&(p_{1}-p_{-1})[\epsilon u_{t}v_{t}+\lambda(u_{t}^{2}{\cal
B}_{t}^{*}-v_{t}^{2}{\cal B}_{t})]+\nonumber \\ \nonumber\\
&-&2i\lambda^{2}(u_{t}^{*}v_{t}x_{t}-u_{t}v_{t}^{*}y_{t})I_{3}^{*} -
2i\lambda^{2}(u_{t}v_{t}^{*}x_{t}^{*}-u_{t}^{*}v_{t}y_{t}^{*})
I_{2}+\nonumber\\
\nonumber\\
&+&i\lambda^{2}(u_{t}^{2}y_{t}+v_{t}^{2}x_{t})I_{1}^{*}+i\lambda^{2}
(u_{t}^{2}x_{t}^{*}+v_{t}^{2}y_{t}^{*})I_{1}\\ \nonumber\\
\label{50}
i(1+2\nu)(\dot x_{t}^{*}y_{t}^{*}-x_{t}^{*}\dot y_{t}^{*})&=&
-(1+2\nu)2\omega x_{t}^{*}y_{t}^{*} -
2i\lambda^{2}(u_{t}^{*}v_{t}y_{t}^{*} - u_{t}v_{t}^{*}x_{t}^{*})
I_{1}^{*}+\nonumber\\ \nonumber\\
&+&2i\lambda^{2}({u_{t}^{*}}^{2}y_{t}^{*} + {v_{t}^{*}}^{2}x_{t}^{*})
I_{3}^{*} + 2i\lambda^{2}(u_{t}^{2}x_{t}^{*} +
v_{t}^{2}y_{t}^{*})I_{2}^{*}\\ \nonumber
\end{eqnarray}

\noindent where

\begin{eqnarray}
\label{51}
I_{1}&=&\int_{0}^{t}dt'
(u_{t'}v_{t'}^{*}y_{t'}-u_{t'}^{*}v_{t'}x_{t'})[p_{1}(1-p_{1}) +
p_{-1}(1-p_{-1})]_{t'}e^{i\varphi(t,t')}\\ \nonumber\\
\label{52}
I_{2}&=&\int_{0}^{t}dt'(u_{t'}^{2}y_{t'}+v_{t'}^{2}x_{t'})
[p_{-1}(1-p_{1})+(p_{-1}-p_{1})\nu]_{t'} e^{-i(2\xi(t,t') +
\varphi(t,t'))}\\ \nonumber\\
\label{53}
I_{3}&=&\int_{0}^{t}dt'({u_{t'}^{*}}^{2}x_{t'}+{v_{t'}^{*}}^{2}
y_{t'})[p_{1}(1-p_{-1}) + (p_{1}-p_{-1})\nu]_{t'} e^{i(2\xi(t,t') -
\varphi(t,t'))}\\ \nonumber
\end{eqnarray}

Using the phase equation (\ref{44}) one can write the integral $I_{1}$
as

\begin{equation}
\label{54}
I_{1}=e^{-I_{\beta}}I_{e}^{(1)}(t)
\end{equation}

\noindent where

\begin{eqnarray}
I_{\beta}(t)&=&\int_{0}^{t}dt' F_{\beta}(t') \\ \nonumber\\
I_{e}^{(1)}(t)&=&\int_{0}^{t}dt'
(u_{t'}v_{t'}^{*}y_{t'}-u_{t'}^{*}v_{t'}x_{t'})[
p_{1}(1-p_{1})+p_{-1}(1-p_{-1})]_{t'}e^{i\varphi(t',0)}\\ \nonumber\\
F_{\beta}&=&i(\dot x_{t}^{*}x_{t}-\dot
y_{t}^{*}y_{t})+\omega(|x_{t}|^{2}+|y_{t}|^{2})\\ \nonumber
\end{eqnarray}

To solve the equation (\ref{54}) we write differential equation for
the integrals $I_{\beta}$ and $I_{e}^{(1)}$

\begin{eqnarray}
\label{55}
\dot I_{\beta}&=&i(\dot x_{t}^{*}x_{t}-\dot
y_{t}^{*}y_{t})+\omega(|x_{t}|^{2}+|y_{t}|^{2})\\ \nonumber\\
\label{56}
\dot I_{e}^{(1)}&=&(u_{t}v_{t}^{*}y_{t}-u_{t}^{*}v_{t}x_{t})[
p_{1}(1-p_{1})+p_{-1}(1-p_{-1})]_{t}e^{-I_{\beta}}\\ \nonumber
\end{eqnarray}

Using the other phase equation we obtain, from the integrals $I_{2}$ and
$I_{3}$, the following differential equations

\begin{eqnarray}
\label{57}
\dot I_{e}^{(2)}&=&(u_{t}^{2}y_{t}+v_{t}^{2}x_{t})[p_{-1}(1-p_{1}) +
(p_{- 1}-p_{1})\nu]_{t}e^{-i(2I_{\alpha}+I_{\beta})}\\ \nonumber\\
\label{58}
\dot I_{e}^{(3)}&=&({u_{t}^{*}}^{2}x_{t}+{v_{t}^{*}}^{2}y_{t})
[p_{1}(1-p _{-1}) + (p_{1}-p_{-1})\nu]_{t} e^{i(2I_{\alpha}-I_{\beta})}\\
\nonumber\\
\label{59}
\dot I_{\alpha}&=&i(\dot u_{t}^{*}u_{t}+\dot v_{t}^{*}v_{t})+\frac
{\epsilon}{2}(|u_{t}|^{2}-|v_{t}|^{2})-\lambda(u_{t}^{*}v_{t}{\cal
B}_{t}+u_{t}v_{t}^{*}{\cal B}_{t}^{*})\\  \nonumber
\end{eqnarray}

Thus, we can integrate the equations (\ref{55}-\ref{56}) and
(\ref{57}-\ref{59}) together with the equations (\ref{47}-\ref{50}).
These equations describe the time evolution of the gaussian variables
in the collisional approximation.

\section{Results and Discussion}

We give below results obtained for two relevant observables associated
with the spin subsystem of the Jaynes-Cummings Hamiltonian: the atomic
inversion, $\langle\sigma_{3}\rangle_{t}$, and the ``intrinsic''
atomic inversion, defined in terms of the projection of $\vec{\sigma}$
along a unit vector pointing along the direction of the spin vector,
$\langle\vec{\sigma}\rangle / |\langle\vec{\sigma}\rangle |$, which
will be denoted by $\langle\sigma_{p}\rangle_{t}$ \cite{marta}. The
latter quantity corresponds in fact to the degree of polarization of
the spin 1/2 subsystem, and hence is directly related to the purity of
the corresponding reduced density. Thus any time dependence of this
quantity implies non-isoentropic behavior of the subsystem
dynamics. In terms of the Bogoliubov parameters, the mean values of
these observables are given by

\[
\langle\sigma_{3}\rangle_{t}=(p_{1}-p_{-1})\cos{2\theta}
\]
\noindent
\[
\langle\sigma_{p}\rangle_{t}=(p_{1}-p_{-1})
\]

\noindent where we use the parametrization

\begin{eqnarray}
u_{t}&=&\cos\theta\nonumber\\\nonumber\\
v_{t}&=&e^{-i\delta}\sin\theta. \nonumber
\end{eqnarray}

A typical example of the results obtained within the present
approximations for the time evolution of these objects is shown in the
figures below. They have been obtained by solving numerically the
dynamical equations (\ref{47}-\ref{50}) and are compared with the
corresponding exact solutions of the model. Figure 1 shows the time
evolution of the atomic inversion $\langle\sigma_{3}\rangle_{t}$ when
the atom is initially prepared in the excited state ($p_{1}(0)=1$,
$p_{-1}(0)=0$) and the field is initially in the coherent state
constructed as the vacuum of the displaced annihilation operator $b$
($x_{t=0}=1$, $y_{t=0}=0$ and $\nu(0)=0$). The mean photon number
${\cal{B}}_{t=0}^2$ is 25 in this example. We use units such that
$\epsilon=\omega=1$ and $\lambda=0.5$ . The evolution of the intrinsic
polarization for the same initial conditions is shown in Figure 2.

The mean-field approximation amounts to ignoring the integrals
$I_{1}$, $I_{2}$ and $I_{3}$, which are set to zero at all times in
this case. This causes the occupation probabilities $\nu$, $p_1$
and $p_{-1}$ to become time-independent, so that there are no
correlations developing between the subsystems. As a consequence,
there is no depolarization and the inversion undergoes Rabi
oscillations with constant amplitude. In the collisional approximation
a modulation of the amplitude of the Rabi oscillations is obtained
from the depolarization resulting from the approximate collisional
corrections to the mean field approximation, included in
Eqs. (\ref{47}) to (\ref{53}), which allow for {\it intersubsystem}
correlations to develop. The effectiveness of the present treatment of
these corrections can be judged by comparing its results with the
exact solution. As can be seen in the figures, the approximate
solution is quantitatively adequate up to times comparable with the
depolarization time.

We may infer from this that the inclusion of the present approximation
of the collision integrals in the dynamical equations is not only
fundamental to generate the qualitative behavior associated with the
decorrelation process related to the initial damping of the Rabi
oscillations, but also successfully describes such effects
quantitatively over a time span covering at least several Rabbi
periods. Improvements of the collisional approximation implemented
here are however needed if one wishes to extend time range of
quantitative reliability.

Finally, we briefly comment on the extension of the present treatment
to 3+1 dimensions in a field-theoretical context (the scalar plasma
\cite{Kal}). In general, an entirely analogous procedure will apply
also to this case, if only one expands the field operators in terms of
the eigenfunctions of the corresponding extended one-fermion and
one-boson densities. This is in general a time-dependent basis whose
evolution will be determined by generalized forms of Eqs. (\ref{49})
and (\ref{50}). The case of a spatially uniform system is particularly
simple, since in this case the extended densities are diagonal in a
plane-wave representation. It can be handled therefore in complete
analogy with the 0+1 dimensional case if one uses a momentum expansion
of the field operators \cite{Lin}. As expected, the usual infinities
of quantum field theory show up and can be absorbed by introducing
appropriate counterterms. Work in this line is under way.

\vskip 0.5cm  
\noindent
{\bf Acknowledgements}
\vskip 0.5cm
This work was supported by Conselho Nacional de
Desenvolvimento Cient\'{\i}fico e Tecnol\'ogico (CNPq), Brazil.

\newpage
\topskip 0.0cm

\newpage
\centerline{\Large\bf Figure Captions}
\vskip 1.0cm 
Fig.1. Time evolution of the atomic inversion with
$\lambda=0.5$. Initial conditions: $\nu=0$, $x_{t}=1$, $y_{t}=0$, $|{\cal
B}_{t}|=5$ (coherent state); $p_{1}=1$, $p_{-1}=0$. Full line: exact
solution; dot-dashed line: collisional approximation; dashed line:
mean-field approximation.

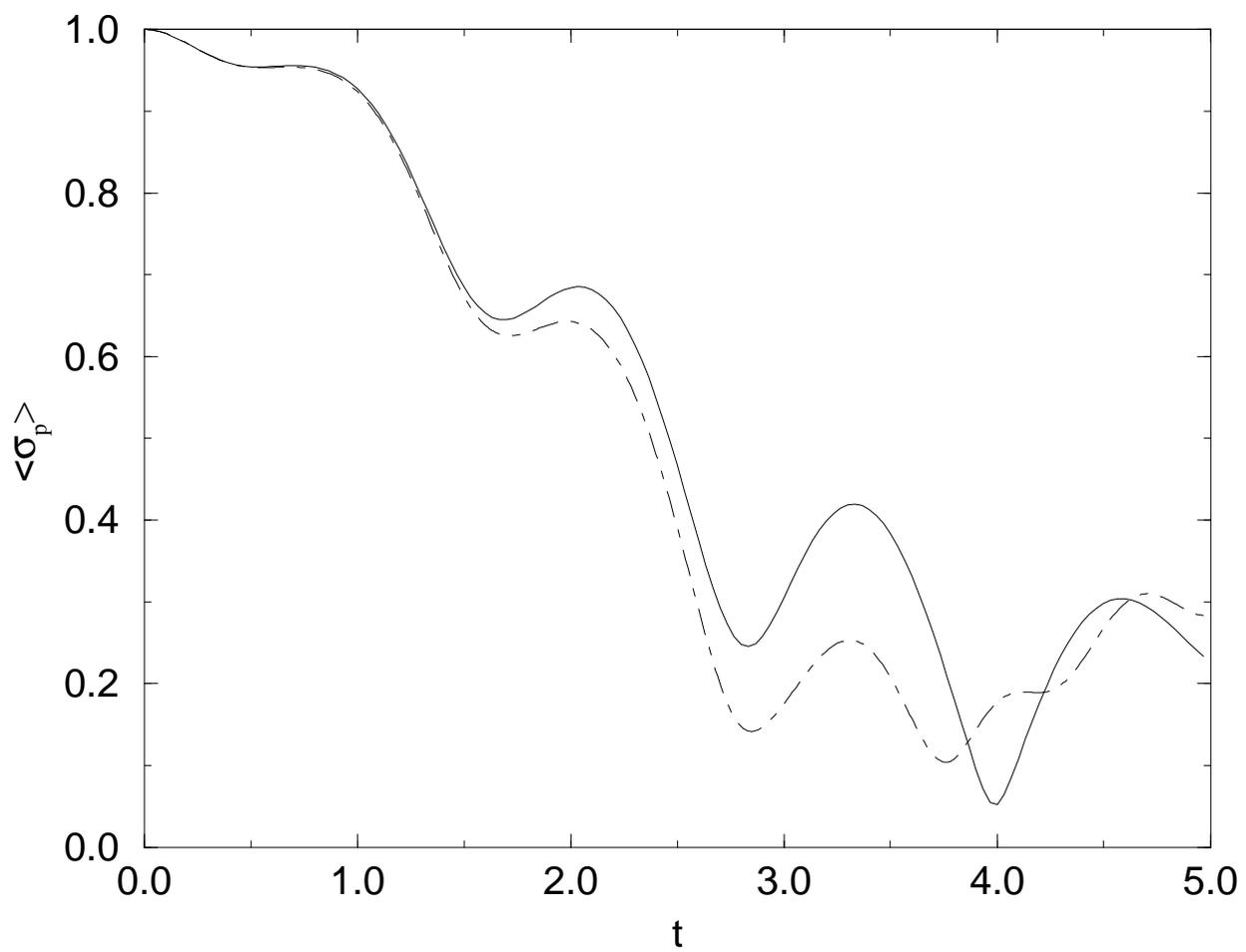
Fig.2. Time evolution of the intrinsic polarization. The initial
conditions are the same as Fig. 1. Full line: exact solution;
dot-dashed line: collisional approximation.

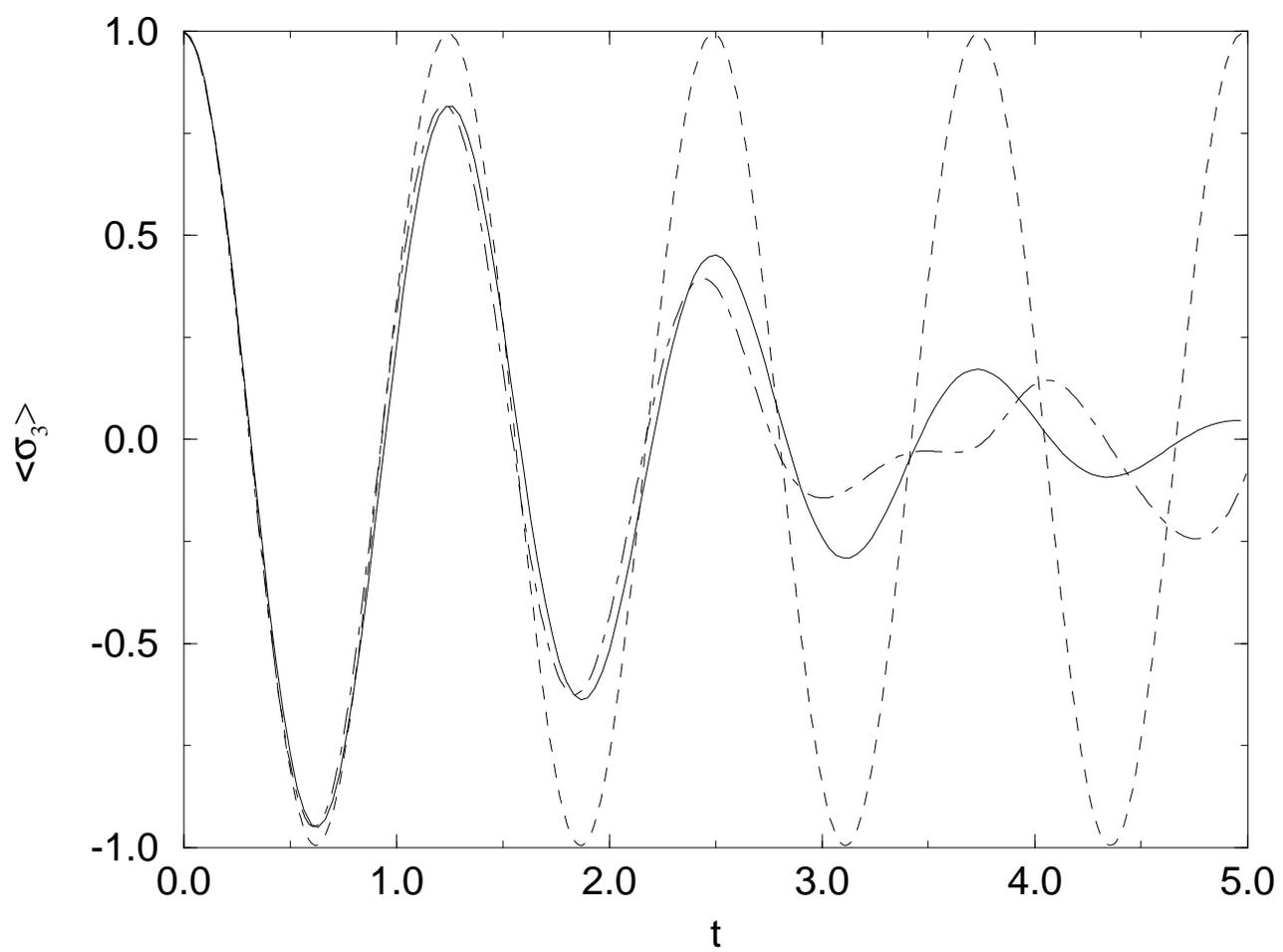
\end{document}